\documentclass[prd,onecolumn,nopacs,nofootinbib]{revtex4}

\usepackage{amsmath}
\usepackage{graphicx}
\usepackage{amsfonts}
\usepackage{latexsym}
\usepackage{bbold}
\usepackage{wasysym}
\usepackage{calligra}
\usepackage{float}
\usepackage{ulem}
\usepackage{inputenc}
\usepackage{xspace}
\usepackage{url}
\usepackage{epstopdf}
\usepackage{tikz}
\usepackage{verbatim}

\newcommand{\be}{\begin{equation}}
\newcommand{\ee}{\end{equation}}
\newcommand{\beq} {\begin{equation}}
\newcommand{\eeq} {\end{equation}}
\newcommand{\ba}{\begin{eqnarray}}
\newcommand{\ea}{\end{eqnarray}}

\begin{document}
	
\title{Friedmann-like universes with non-metricity}

\author{Damianos Iosifidis}
\email{diosifid@auth.gr}
\affiliation{Institute of Theoretical Physics, Department of Physics, Aristotle University of Thessaloniki, Thessaloniki 54124, Greece}

\author{Ioannis Georgios Vogiatzis}
\email{ivogiatz@physics.auth.gr}
\affiliation{Department of Physics, Aristotle University of Thessaloniki, Thessaloniki 54124, Greece}

\author{Christos G. Tsagas}
\email{tsagas@astro.auth.gr}
\affiliation{Section of Astrophysics, Astronomy and Mechanics, Department of Physics, Aristotle University of Thessaloniki, Thessaloniki 54124, Greece\\ and\\ Clare Hall, University of Cambridge, Herschel Road, Cambridge, CB3 9AL, UK}	


\begin{abstract}
We study  the potential effects of spacetime non-metricity  in cosmology. In the spirit of Einstein-Cartan gravity, but with non-metricity replacing torsion, we consider the Einstein-Hilbert action and assume zero torsion. Adopting certain hyperfluid models, with non-vanishing hypermomentum that can source spacetime non-metricity, we add a matter component into the action and derive the field equations, along with the conservation laws. Applying our formulae to cosmology, we generalize the Friedmann and the Raychaudhuri equations in the presence of non-metricity. Our results show that, in a number of cases, non-metricity can mimic the effects of matter with unconventional equation of state. For instance, specific types of hypermomentum are found to act as an effective stiff fluid, thus opening the possibility that non-metricity could have played a significant role in the early stages of the universe's evolution. Alternative forms of hypermomentum can dominate the universal dynamics at late times. In either case, the equilibrium moment depends on the initial conditions and it is determined by a simple relation between the matter component and the hyperfluid.
\end{abstract}
	
\maketitle
	
\allowdisplaybreaks
	
	

	
\section{Introduction}\label{sI}
Non-Riemannian geometries~\cite{eisenhart2012non} offer a natural playground for developing gravitational theories beyond the Riemannian framework of general relativity. The additional geometric structures of such theories come from both torsion and non-metricity, as is the case of the Metric-Affine Gravity (MAG) framework~\cite{hehl1995metric,iosifidis2019metric}. Naturally, such fundamental modifications have a strong impact on the cosmic evolution and alter the Friedmann equations in a non-trivial manner~\cite{puetzfeld2005prospects}. For instance, in~\cite{kranas2019friedmann,pereira2019acceleration,Guimaraes:2020drj} and also in~\cite{Iosifidis:2021iuw}, it was shown that torsion can drastically alter the cosmological evolution and provide intriguing new results. From the geometrical point of view, torsion reflects the fact that the parallel transport of two vectors along each other does not form a closed parallelogram. As for non-metricity, it manifests the inability of the metric to be covariantly conserved, which in turn implies that lengths and angles are not preserved during the parallel transport of vectors~\cite{iosifidis2019metric}. Possible experimental effects/manifestations of torsion and of non-metricity have been studied by Kostelecky~\cite{kostelecky2008constraints}, while other potential signatures of non-metricity have been investigated in~\cite{delhom2018observable,bahamonde2021observational}.

These non-Riemannian degrees of freedom, namely torsion and non-metricity, are sourced by the hypermomentum tensor of the matter~\cite{hehl1976hypermomentum}. The latter is formally defined as the variational derivative of the matter action with respect to the connection and encapsulates the microstructure properties of the matter fields, that is \textit{dilation}, \textit{shear} and \textit{spin}. Continua with non-vanishing hypermomentum tensor have been constructed in \cite{obukhov1993hyperfluid, obukhov1996model,babourova1995variational} and the development of the Perfect Hyperfluid structure has been formulated in~\cite{Iosifidis:2020gth}, with its generalization given in~\cite{iosifidis2021perfect}. The latter extends the classical perfect fluid notion of general relativity (GR) to account for the microspcopic properties of the matter. This model can be used to study the cosmological aspects of torsion and non-metricity, since it excites all their degrees of freedom in a homogeneous and isotropic Friedmann-like universes~\cite{Iosifidis:2020gth}. The role of torsion in Friedmann-like universes has been studied by several authors (e.g.~see~\cite{kranas2019friedmann,Iosifidis:2021iuw, barrow2019friedmann,pereira2019acceleration}). Here, instead, we focus on non-metricity and study its role in homogeneous and isotropic Friedmann-like universes.\footnote{The cosmological aspects of the general quadratic metric-affine gravity with torsion and non-metricity have been recently studied in~\cite{iosifidis2022cosmology}.} We find that, depending on the form of the hypermomentum tensor, the effects of spacetime non-metricty can mimic those of matter with different equations of state. When the hyperfluid has only a dilation component, for example, it behaves like stiff matter and it can dominate the early evolution of the universe. When dealing with purely shear hypermomentum, on the other hand, the effects of non-metricity are identical to those of a perfect fluid with an arbitrary barotropic index and can dominate at late times as well. Whether the moment of equilibrium (between the matter component and the hyperfluid) lies in the past or in the future, is determined by a simple relation that depends on the initial conditions.

The paper is organised as follows. In section~II we develop the  framework, including the necessary definitions and the basic geometric notions of non-metricity, the matter conservation laws and the field equations of the theory. Then, in section~III, we outline the key cosmological aspects of Friedmann-like universes with non-metricity and the sources that produce them.
Section~IV is primarily devoted to the investigation of the hyperfluid model proposed in~\cite{obukhov1996model}. In section~V we adopt the most general perfect cosmological (hypermomentum-preserving) hyperfluid and demonstrate how it excites all the non-metricity degrees of freedom. There, we also derive the modified Friedmann and Raychaudhuri equations, together with the associated conservation laws. These are accompanied by a number of exact cosmological solutions and their physical interpretation. Finally, we conclude with a discussion of our results and an outline of possible future extensions.

\section{The theoretical framework}
The fundamental novelty of non-metricity is that, in direct contrast to Riemannian geometry, the metric tensor is no longer covariantly constant. Put another way, the covariant derivative of the metric is not necessarily zero. This introduces a number of unconventional degrees of freedom to any theoretical model of the universe.

\subsection{The non-metricity tensor}\label{ssN-Mt}
Relaxing the metricity constraint allows magnitudes and angles to change during the parallel transport of vectors and tensors. These changes reflect the fact that the metric tensor is not covariantly conserved, and its failure to remain so is measured by the non-metricity tensor
\beq
Q_{\lambda\mu\nu}= -\nabla_{\lambda}g_{\mu\nu}\,,  \label{Q}
\eeq
with $Q_{\lambda\mu\nu}=Q_{\lambda(\mu\nu)}$ by construction. The symmetries of $Q_{\lambda\mu\nu}$ ensure that one can construct a pair of independent non-metricity vectors, namely
\beq
Q_{\lambda}= Q_{\lambda\mu\nu}g^{\mu\nu} \hspace{15mm} {\rm and} \hspace{15mm} q_{\nu}= Q_{\lambda\mu\nu}g^{\lambda\mu}\,,  \label{Qvecs}
\eeq
the former of which is usually referred to as the Weyl vector.

In addition to non-metricity, spacetime torsion is also present in a general non-Riemannian space. The torsion tensor is defined as the antisymmetric part of the connection

\beq
S_{\kappa\lambda}^{\;\;\;\;\mu}=\Gamma^\mu_{\;\;\;[\kappa\lambda]} \;.\label{torsiondefinition}
\eeq
out of which one can construct the torsion vector
\beq
S_{\mu}=S_{\mu\nu}^{\;\;\;\;\nu}\,.
\eeq
Let us note that, with the exception of this section, in the rest of this manuscript the connection will be treated as symmetric and therefore the space will be assumed torsion free.

\subsection{Connection and curvature}\label{ssCC}
In the general metric-affine framework, the connection and the metric are regarded as independent variables. Thus, contrary to GR, the connection is no longer expressed as a function of the metric (and of its derivatives) alone. More specifically, in a torsion-free space, the symmetric connection is given by
\beq
\Gamma^\lambda_{\;\;\;\mu\nu}= \tilde{\Gamma}^\lambda_{\;\;\;\mu\nu}+ \frac{1}{2}\, g^{\rho\lambda}\left(Q_{\mu\nu\rho}+Q_{\nu\rho\mu} -Q_{\rho\mu\nu}\right)\,.  \label{ConnectionDecomp}
\eeq
The first term on the right-hand side of the above is the Levi-Civita connection, with
\beq
\tilde{\Gamma}^\lambda_{\;\;\;\mu\nu}=\frac{1}{2}\, g^{\rho\lambda}\left( \partial_\mu g_{\nu\rho}+\partial_\nu g_{\rho\mu}-\partial_\rho g_{\mu\nu}\right)\,,
\eeq
while the rest form the so-called ``distortion tensor''\footnote{In spaces where only non-metricity is present, the distortion tensor is sometimes referred to as ``deflection''.}
\beq
N^\lambda_{\;\;\;\mu\nu}= \Gamma^\lambda_{\;\;\;\mu\nu}-\tilde{\Gamma}^\lambda_{\;\;\;\mu\nu}= \frac{1}{2}\,g^{\rho\lambda} \left(Q_{\mu\nu\rho}+Q_{\nu\rho\mu} -Q_{\rho\mu\nu}\right)\,.  \label{distortiontensordefinition}
\eeq
Therefore, a general affine connection decomposes into its Levi-Civita component and a non-metricity contribution. As a result, any quantity that depends on the connection splits into a Riemannian and a non-Riemannian part.

In the presence of non-metricity, spacetime curvature is determined by the associated generalised Riemann tensor. The latter satisfies an expression formally identical to that of its Riemannian analogue, namely
\beq
R^\mu_{\;\;\;\nu\alpha\beta}= 2\partial_{[\alpha}\Gamma^\mu_{\;\;\;|\nu|\beta]}+ 2\Gamma^\mu_{\;\;\;\rho[\alpha}\Gamma^\rho_{\;\;\;|\nu|\beta]}\,. \label{Riemmantensordefinition}
\eeq
Nevertheless, $R^\mu_{\;\;\;\nu\alpha\beta}$ lacks some of the symmetries we are accustomed with from standard GR. As a result, in non-Riemannian manifolds, there are three independent contractions of the Riemann tensor. The first corresponds to the familiar Ricci tensor defined as
\beq
R_{\nu\beta}=R^\mu_{\;\;\;\nu\mu\beta}= 2\partial_{[\mu}\Gamma^\mu_{\;\;\;|\nu|\beta]}+ 2\Gamma^\mu_{\;\;\;\rho[\mu}\Gamma^\rho_{\;\;\;|\nu|\beta]}\,,
\eeq
which (in contrast to GR) is not necessarily symmetric. The other two contractions are known as the ``homothetic curvature tensor'' and the ``co-Ricci tensor''. These are given by\footnote{The symmetries of the conventional curvature tensor ensure that the homothetic curvature tensor vanishes in Riemannian spaces, while the co-Ricci tensor is the opposite of the Ricci tensor.}
\beq
\hat{R}_{\alpha\beta}=R^\mu_{\;\;\;\mu\alpha\beta} \hspace{15mm} {\rm and} \hspace{15mm} \check{R}^\mu_{\;\;\;\beta}= g^{\nu\alpha}R^\mu_{\;\;\;\nu\alpha\beta}\,,
\eeq
respectively. Finally, the trace of the above leads to the uniquely defined Ricci scalar
\beq
R=R^\beta_{\;\;\;\beta}=g^{\nu\beta}R_{\nu\beta}\,, \hspace{10mm} {\rm with} \hspace{10mm} \hat{R}_{\mu\nu}g^{\mu\nu}\equiv 0\,.
\eeq
It goes without saying that all these tensors and scalars also decompose into their Riemannian and non-metricity parts (see section~\ref{sFUN-M} below).

\subsection{Sources and conservation laws}
In metric affine gravity the action is a functional of the metric, of the independent affine connection and of the various matter fields. In other words,
\beq
S[g,\Gamma,\phi]= S_G[g,\Gamma]+ S_M[g,\Gamma,\phi]\,,
\eeq
where $S_G$ is the gravitational and $S_M$ is the matter component of the action.\footnote{Gravitational theories where the matter part of the action does not explicitly depend on the connection are often called Palatini theories.} The tensors associated to the matter are the usual (metrical) energy-momentum tensor
\beq
T^{\alpha\beta}=\frac{2}{\sqrt{-g}}\frac{\delta(\sqrt{-g}\, \mathcal{L}_{M})}{\delta g_{\alpha\beta}}
\eeq
and (since matter couples to the connection) the hypermomentum tensor. The latter encapsulates the microscopic characteristics of the matter (i.e.~spin, dilation and shear) and it is given by~\cite{hehl1976hypermomentum}
\beq
\Delta_{\lambda}^{\;\;\;\mu\nu}= -\frac{2}{\sqrt{-g}} \frac{\delta(\sqrt{-g}\,\mathcal{L}_{M})} {\delta\Gamma^{\lambda}_{\;\;\;\mu\nu}}\,.
\eeq
Working in the equivalent exterior calculus formalism and using the vielbien $e_\mu^{\;\;\;c}$, we can also define the canonical energy-momentum tensor by means of
\beq
t^{\mu}_{\;\;c}= \frac{1}{\sqrt{-g}}\,\frac{\delta S_{M}}{\delta e_{\mu}^{\;\; c}}\,.
\eeq

The metrical and the canonical energy-momentum tensors are not independent, but they are related and obey the conservation laws  \cite{obukhov2013conservation,Iosifidis:2020gth}
\beq
t^{\mu}_{\;\;\lambda}= T^{\mu}_{\;\;\lambda}- \frac{1}{2\sqrt{-g}}\, \hat{\nabla}_{\nu}(\sqrt{-g}\,\Delta_{\lambda}^{\;\;\mu\nu}) \label{cc1}
\eeq
and
\beq \frac{1}{\sqrt{-g}}\, \hat{\nabla}_{\mu}(\sqrt{-g}\,t^{\mu}_{\;\;\alpha})= -\frac{1}{2}\, \Delta^{\lambda\mu\nu}R_{\lambda\mu\nu\alpha}+ \frac{1}{2}\,Q_{\alpha\mu\nu}T^{\mu\nu}+2 S_{\alpha\mu\nu}t^{\mu\nu}\,,  \label{cc2}
\eeq
with $S_{\alpha\mu\nu}$ and $S_\nu$ representing the torsion tensor and torsion vector respectively. Also, the operator $\hat{\nabla}_{\nu}$ seen in both of the above is defined as
\beq
\hat{\nabla}_{\nu}:= 2S_{\nu}-\nabla_{\nu}\,.
\eeq
Note that our conservation laws (\ref{cc1}) and (\ref{cc2}) follow directly from the diffeomorphic invariance of the matter action and also from the GL invariance that the theory exhibits when written in the language of differential forms.

Let us now switch the torsion off and also adopt the scheme of~\cite{Iosifidis:2020gth}, where the metrical and the canonical energy-momentum tensors coincide. This assumption does not always hold, since the metrical energy-momentum tensor is symmetric and the canonical is generally not. Nevertheless, in homogeneous and isotropic spacetimes (like those considered here), both tensors are symmetric and the above assumption holds. In addition, allowing the metrical and the canonical energy-momentum tensors to differ, leads to a hypermomentum tensor that is not conserved and to considerably more involved equations. We shall therefore confine to hypermomentum-conserving hyperfluids, with
\beq
\nabla_{\nu}(\sqrt{-g}\,\Delta_{\lambda}^{\;\;\mu\nu})= 0\ \label{QlawHM}
\eeq
and
\beq \frac{1}{\sqrt{-g}}\,\nabla_{\mu}(\sqrt{-g}\,T^{\mu}_{\;\;\alpha})= \frac{1}{2}\,\Delta^{\lambda\mu\nu}R_{\lambda\mu\nu\alpha}- \frac{1}{2}\,Q_{\alpha\mu\nu}T^{\mu\nu}\,,  \label{QlawEM}
\eeq
throughout the rest of this manuscript.

\subsection{The field equations}\label{ssFEs}
Our aim is to study the cosmological implications of a gravitational theory that is analogous to the classical Einstein-Cartan gravity, but with non-metricity replacing torsion. To this end, we shall consider the action
\beq
S[g,\Gamma,\varphi]=\frac{1}{2\kappa}\int d^{n}x \sqrt{-g} R+ S_{M}\,,  \label{action}
\eeq
where the first term on the right-hand side is the usual Einstein-Hilbert action, $S_{M}$ denotes the matter sector and $\varphi$ represents the matter fields that produce spacetime non-metricity. Since the latter is at the centre of our study, we will assume a symmetric (i.e.~torsion-free) connection from now on. Then, variation with respect to the metric and the independent connection, yields the field equations
\beq
R_{(\mu\nu)}- \frac{1}{2}\,Rg_{\mu\nu}= \kappa T_{\mu\nu} \hspace{15mm} {\rm and} \hspace{15mm} P_{\lambda}^{\;\;(\mu\nu)}= \kappa\Delta_{\lambda}{}^{(\mu\nu)}\,,  \label{fe2}
\eeq
with
\beq
P_{\lambda}{}^{(\mu\nu)}= \frac{1}{2}\,Q_{\lambda}g^{\mu\nu}-Q_{\lambda}{}^{\mu\nu}+ \left(q^{(\mu}-\frac{1}{2}\,Q^{(\mu}\right) \delta_{\lambda}{}^{\nu)}\,,  \label{Palatini}
\eeq
giving the symmetrised (torsion-free) Palatini tensor. The first set of formulae represents the modified Einstein equations in the presence of non-metricity, while the second shows how the hypermomentum sources non-metricity. In what follows, we will apply the above to a homogeneous and isotropic Friedmann-like universe.

\section{Cosmology with non-metricity}\label{sCN-M}
The Friedmann models are characterised by their spatial homogeneity and isotropy. Non-metricity introduces new degrees of freedom to all cosmological scenarios and broadens their phenomenology. This is also true for the Friedmann universes, although their high symmetry demands that the matter content has the perfect-fluid form.

\subsection{Geometrical aspects of Friedmann-like universes
non-metricity}
Let us consider a spatially flat (i.e.~with $K=0$, where $K$ is the 3-curvature index) Friedmann-Lemaitre-Robertson-Walker (FLRW) spacetime, satisfying the familiar Robertson-Walker line element
\beq\label{RWle}
ds^{2}=-dt^{2}+a^{2}\delta_{ij}dx^{i}dx^{j} \,,
\eeq
where $a=a(t)$ is the cosmological scale factor and $i,j=1,2,3$. Introducing the normalized $4$-velocity field $u_\mu$, with co-moving coordinates $u^\mu= \delta^\mu{}_0=(1,0,0,0)$ and $u_\mu u^\mu=-1$, the associated temporal-derivative operator is\footnote{Evaluating the temporal derivative with respect to the Levi-Civita connection for a scalar yields the same result.}
\beq\label{tempdev}
\dot{}=u^\alpha \nabla_\alpha \,.
\eeq
In addition, the symmetric tensor
\beq\label{projop}
h_{\mu \nu}= g_{\mu \nu} + u_\mu u_\nu \,,
\eeq
acts as the metric of the spatial hypersurfaces. By construction, the above projects into the 3-space orthogonal to $u^\mu$ (since $h_{\mu\nu}u^\nu=0$) and also satisfies the constraints $h_\mu{}^\mu=3$ and $h_{\mu\alpha}h^{\alpha}{}_\nu=h_{\mu\nu}$.
On using the $u_\mu$ and the $h_{\mu\nu}$ fields, one achieves an 1+3 decomposition of the spacetime into time and 3-D space.

In non-Riemannian Friedmann-like spacetimes, torsion is described by two and non-metricity by three degrees of freedom, as it was respectively shown in \cite{tsamparlis1979cosmological,minkevich1998isotropic}. Covariantly written, the non-metricity tensor reads \cite{Iosifidis:2020gth}
\begin{equation}\label{nonmetFLRW}
Q_{\alpha \mu \nu}  = Au_\alpha h_{\mu \nu} + Bh_{\alpha(\mu} u_{\nu)} + Cu_\alpha u_\mu u_\nu \,,
\end{equation}
the contractions of which lead to the non-metricity vectors
\begin{equation}\label{nmtracesFLRW}
Q_\mu = (3A-C)u_\mu \hspace{15mm} {\rm and} \hspace{15mm}
q_\mu= \left(\frac{3}{2}\,B-C\right)u_\mu\,.
\end{equation}
The functions $A=A(t)$, $B=B(t)$ and $C=C(t)$ monitor the cosmological effects of non-metricity in Friedmann-like universes. Also, together with the scale factor, the above describe non-metric FLRW-like spacetimes. For instance, along with the FLRW metric, expressions (\ref{nonmetFLRW}) and (\ref{nmtracesFLRW}) can be used to calculate the Riemann tensor and its contractions (by means of Eq.~(\ref{ConnectionDecomp}) -- see Appendix~\ref{ApA}).

\subsection{The perfect cosmological hyperfluid}\label{ssUH}
In order to probe the implications of non-metricity for cosmology, we will consider an FLRW-type universe filled with a perfect hyperfluid~\cite{Iosifidis:2020gth}. The latter generalises the familiar perfec fluid by accounting for the non-Riemannian degrees of freedom. Then, relative to a family of observers with 4-velocity $u_{\mu}$, the energy-momentum tensor of the matter reads
\beq
T_{\mu\nu}= \rho u_{\mu}u_{\nu}+ ph_{\mu\nu}\,,
\eeq
with $\rho$ and $p$ representing the energy density and the isotropic pressure of the matter respectively. As we have already discussed, the canonical energy-momentum tensor also takes the above form. In addition, demanding that the hypermomentum tensor obeys the Cosmological Principle, we have~\cite{Iosifidis:2020gth}
\beq
\Delta_{\alpha\mu\nu}=\phi h_{\mu\alpha}u_{\nu}+\chi h_{\nu\alpha}u_{\mu}+\psi u_{\alpha}h_{\mu\nu}+\omega u_{\alpha}u_{\mu} u_{\nu}+\epsilon_{\alpha\mu\nu\rho}u^{\rho}\zeta \,.
\eeq
The above sources both torsion and non-metricity, with the former related to the spin part ($\Delta_{[\alpha\mu]\nu}$). Therefore, in order to isolate the non-metricity effects, we set  $\Delta_{[\alpha\mu]\nu}=0$. This ensures that $\psi=\chi$ and subsequently leads to
 \beq
\Delta_{\alpha\mu\nu}=\phi h_{\mu\alpha}u_{\nu}+2\psi u_{(\alpha}h_{\mu)\nu}+\omega u_{\alpha}u_{\mu} u_{\nu}\,.
\eeq
We have now developed all the necessary formalism needed for the rest of our analysis. Next, in section~\ref{sFUN-M}, we will start with the cosmological implications of the Obukhov hyperfluid~\cite{obukhov1996model}. Then, in section~\ref{GNM}, we turn our attention to the generalised non-metricity tensor (\ref{nonmetFLRW}) developed in~\cite{Iosifidis:2020gth}.

\section{Cosmology with Obukhov's hyperfluid}\label{sFUN-M}
In Obukhov's model the canonical energy-momentum tensor coincides with the metrical, while the hypermomentum tensor has the special (density)x(current) form~\cite{obukhov1996model}. Here, we will further constrain this hypermomentum form, in order to comply with the Cosmological Principle.

\subsection{Energy-momentum and hypermomentum}
As mentioned above, the energy-momentum tensor has the familiar perfect fluid form, with $T_{\mu\nu}=\rho u_{\mu} u_{\nu}+p h_{\mu\nu}$. At the same time, the hypermomentum tensor reads
\beq
\Delta_{\lambda}^{\;\;\;\mu\nu}=J_{\lambda}^{\;\;\mu}u^{\nu}\,,
\eeq
where $J_{\mu\nu}$ is the hypermomentum density of the hyperfluid. Given that the antisymmetric part ($J_{[\mu\nu]}$) is related to spin and torsion~\cite{hehl1995metric}, hereafter $J_{\mu\nu}$ will be assumed symmetric. The simplest assumption is that the hypermomentum density is proportional to the projection tensor $h_{\mu\nu}$.\footnote{This is not the only possibility and we will discuss the most general case in the next section.} Then, due to the high symmetry of the FLRW-type host, there exists a time-dependent function $\phi=\phi(t)$ so that
\beq
J_{\mu\nu}=\phi h_{\mu\nu}  \label{J1}
\eeq
and $\Delta_{\alpha\mu\nu}=\phi h_{\alpha\mu}u_{\nu}\, \label{D1}$.
As we will see next, the scalar field $\phi$ completely determines the non-metricity efects.

\subsection{Non-metricity and curvature}
Non-metricity and hypermomentum are related through the hypermomentum field equations. Substituting the hypermomentum form seen in (\ref{D1}) above into Eq.~(\ref{fe2}), leads to the non-metricity tensor
\beq
Q_{\alpha\mu\nu}= -\frac{1}{2}\,\kappa\phi u_{\alpha} (h_{\mu\nu}+u_{\mu}u_{\nu})\,, \label{Qh}
\eeq
the contractions of which provide the associated non-metricity vectors
\beq
Q_{\mu}= -\kappa\phi u_{\mu} \hspace{15mm} {\rm and} \hspace{15mm} q_{\mu}= \frac{1}{2}\,\kappa\phi u_{\mu}\,. \label{Qts}
\eeq

Using the above, one can calculate the corresponding curvature tensors. The Ricci scalar, in particular, decomposes into its Riemannian and non-Riemannian parts as
\beq
R= \tilde{R}+ \frac{1}{4}\,Q_{\alpha\mu\nu}Q^{\alpha\mu\nu}- \frac{1}{2}\,Q_{\alpha\mu\nu}Q^{\mu\nu\alpha}- \frac{1}{4}\,Q_{\mu}Q^{\mu}+ \frac{1}{2}\,Q_{\mu}q^{\mu}+ \tilde{\nabla}_{\mu}(q^{\mu}-Q^{\mu})\,.
\eeq
Note that the tilded variables are evaluated with respect to the Christoffel connection, which makes $\tilde{R}$ the Riemannian Ricci scalar. Then, employing (\ref{Qh}) and (\ref{Qts}), one arrives at
\beq
R= \tilde{R}+ \frac{3}{8}\,\kappa^{2}\phi^{2}+ \frac{3}{2}\,\kappa\frac{1}{a^{3}}\,\partial_{t}(a^{3}\phi)\,, \label{Ric}
\eeq
where
\beq
\tilde{R}= 6\left[\frac{\ddot{a}}{a} +\left(\frac{\dot{a}}{a}\right)^{2}\right]\,,
\eeq
is the Riemannian Ricci scalar. Proceeding in a similar manner one can compute the Ricci tensor in the presence of spacetime non-metricity. In particular, the timelike component of the latter is found to be
\beq
R_{00}= \tilde{R}_{00}- \frac{3}{2}\,A^{2}- \frac{3}{4}\kappa\frac{1}{a^{3}}\partial_{t}\Big( a^{3}\phi \Big)\,,
\eeq
with
\beq
\tilde{R}_{00}=-3\frac{\ddot{a}}{a}\,,
\eeq
representing the Riemannian analogue. Likewise, for the spacelike part of the Ricci tensor, we obtain
\beq
R_{ij}= \tilde{R}_{ij}+ \frac{1}{4}\,\kappa g_{ij} \left[\frac{1}{a^{3}}\,\partial_{t}\left(a^{3}\phi\right)\right]\,,
\eeq
the trace of which ($R=g^{00}R_{00}+g^{ij}R_{ij}$) leads to the Ricci scalar given in Eq.~(\ref{Ric}) previously. Substituting the above post-Riemannian expansions into the metric field equations, one arrives at the generalized Friedmann equations in the presence of non-metric degrees of freedom described by $(\ref{Qh})$.

\subsection{Modified Friedmann equations and conservation laws}
Taking the timelike ($00$) and the spatial ($ij$, with $i,j=1,2,3$) components of the metric field equations and using the curvature relations of the last section, we obtain the modified Friedmann equations. When spscetime non-metricity is sourced by the Obukhov hyperfluid, these take the form
\beq
H^{2}=\left(\frac{\dot{a}}{a}\right)^2= \frac{1}{3}\,\kappa\rho+ \frac{1}{16}\,\kappa^2\phi^{2}
\hspace{15mm} {\rm and} \hspace{15mm} \frac{\ddot{a}}{a}= -\frac{1}{6}\,\kappa(\rho+3p)- \frac{1}{8}\,\kappa^{2}\phi^{2}- \frac{1}{4}\,\kappa\frac{\partial_t(a^{3}\phi)}{a^{3}}\,, \label{FriedE2}
\eeq
with $H=\dot{a}/a$ representing the Hubble parameter (e.g.~see~\cite{tsagas2008relativistic}). The first of the above, which measures the contribution of matter and non-metricity to the Hubble expansion, also recasts as
\begin{equation}
1= \Omega_\rho+ \Omega_\phi\,,  \label{Omegas}
\end{equation}
where $\Omega_\rho=\kappa\rho/3H^2$ is the familiar density parameter (e.g.~see~\cite{tsagas2008relativistic}) and $\Omega_\phi=\kappa^2\phi^2/16H^2$ is its non-metricity analogue. Accordingly, matter dominates the energy density of the universe and drives its expansion, when $\Omega_\rho\gg\Omega_\phi$. In the opposite case, the universe is dominated by spacetime non-metricity.

Turning to the conservation laws, we point out that, in FLRW-type spacetimes with hypermomentum of the form introduced in section~\ref{D1}, one can easily show that
\beq
\Delta^{\alpha\beta\gamma}R_{\alpha\beta\gamma\nu}\equiv 0\,.
\eeq
Then, the conservation law of the associated energy-momentum tensor (see Eq.~$(\ref{QlawEM})$ earlier) reduces to the familiar continuity equation
\beq
\dot{\rho}+ 3H(\rho+p)= 0\,,
\eeq
of a standard perfect fluid~\cite{tsagas2008relativistic}. The corresponding conservation law of  the hypermomentum (see expression ($\ref{QlawHM}$)) provides the evolution formula of $\phi$. By employing the identity
\beq
\nabla_{\lambda}(\sqrt{-g}\phi u^{\lambda})=\partial_{\lambda}(\sqrt{-g}\phi u^{\lambda}),
\eeq
which holds for vanishing torsion, the hypermomentum conservation law reads
\beq
\dot{\phi}+ 3H\phi= 0\,,  \label{conlawA}
\eeq
ensuring the $\phi\propto a^{-3}$. In other words, the non-metricity scalar decays with the expansion at the rate of pressure free ``dust''. Finally, substituting the evolution law of $\phi$ back into Eq.~(\ref{FriedE2}), we arrive at
\beq
\frac{\ddot{a}}{a}= -\frac{1}{6}\,\kappa(\rho+3p)- \frac{1}{8}\,\kappa^{2}\phi_0^2\left(\frac{a_0}{a}\right)^6\,, \label{rayobukov}
\eeq
where the zero suffix corresponds to a chosen moment in time. The above is the modified Raychaudhuri equation, monitoring the acceleration/deceleration of the expansion of a Friedmann-like universe in the presence of Obukhov-type non-metricity. In particular, negative terms on the right-hand side of the above decelerate the expansion, while positive ones tend to accelerate. Alternatively, one can say that the negative terms assist gravitational collapse, whereas the positive ones tend to resist contraction.

Before closing this section, we should note that the non-metricity contribution to both Eqs.~(\ref{FriedE2}a) and (\ref{rayobukov}) is proportional to $a^{-6}$, which means that its effects resemble those of a ``stiff'' fluid with $p=\rho$ (in agreement with~\cite{PhysRevD.56.7769}). Consequently, the Obukhov non-metricity can play a significant role only at early times. This is not unexpected, since the post-Riemannian effects typically dominate at the early stages of the universe~\cite{2005}. Also, following the modified Friedmann equation -- see (\ref{FriedE2}a), non-metricity contributes positively to the total energy density of the universe and it cannot act as an effective ``ghost'' fluid. Finally, according to the modified Raychaudhuri equation -- see (\ref{rayobukov}), the Obukhov non-metricity always tends to decelerate the expansion (or alternatively accelerate the collapse).

\subsection{Generalized unconstrained hyperfluid}
In an FLRW-like universe with non-metrictiy, the hypermmentum density of the hyperfluid ($J_{\mu\nu}$) is generally spanned by two time functions. More specifically, expression (\ref{J1}) generalises to
\beq
J_{\mu\nu}=\phi h_{\mu\nu}+\psi u_{\mu}u_{\nu}\,,
\eeq
with $\psi=\psi(t)$ as well. Then, the corresponding hypermomentum tensor acquires the general form
\beq
\Delta_{\alpha\mu\nu}= (\phi h_{\alpha\mu}+\psi u_{\alpha}u_{\mu})u_{\nu}\,.  \label{D2}
\eeq
In this scenario, the dynamics of the hypermomentum and the effects of non-metricity are determined by $\psi$ as well as $\phi,\psi$. The question then is whether the inclusion of $\psi$ could radically change the results of the previous section. In order to investigate this, we will employ the ``first theorem'' formulated in~\cite{Iosifidis:2018jwu} and solve the connection in terms of the Palatini tensor. The result reads
\begin{equation}
\Gamma^{\lambda}{}_{\mu\nu}= \tilde{\Gamma}^{\lambda}{}_{\mu\nu}+ \frac{1}{2}\,g^{\lambda\alpha} \left(P_{\alpha\mu\nu}-P_{\nu\alpha\mu}-P_{\mu\nu\alpha}\right)+ \frac{1}{2}\,g^{\alpha\lambda}g_{\nu[\mu} \left(P_{\alpha]}-\tilde{P}_{\alpha]}\right)+ \frac{1}{2}\,\delta_{\mu}{}^{\lambda}q_{\nu}\,,  \label{gg1}
\end{equation}
which means that the general distortion tensor (including both Q and S) is given by
\beq
N^{\lambda}{}_{\mu\nu}= \frac{1}{2}\,g^{\lambda\alpha} \left(P_{\alpha\mu\nu}-P_{\nu\alpha\mu}-P_{\mu\nu\alpha}\right)+ \frac{1}{2}\,g^{\alpha\lambda}g_{\nu[\mu} \left(P_{\alpha]}-\tilde{P}_{\alpha]}\right)+ \frac{1}{2}\,\delta_{\mu}{}^{\lambda}q_{\nu}\,.
\eeq
In the case of vanishing torsion, we set $S_{\mu\nu\lambda}=0$ and also symmetrize the above over the indices $\mu$ and $\nu$, since the connection is now symmetric. Therefore, for zero torsion, the distortion tensor simplifies to
\beq
N_{\alpha\mu\nu}= N_{\alpha(\mu\nu)}= \frac{1}{2} \Big(P_{\alpha(\mu\nu)}-P_{(\nu|\alpha|\mu)}-P_{(\mu\nu)\alpha}\Big)+ \frac{1}{2}\,g_{\alpha(\mu}Q_{\nu)}+ \frac{1}{2}\left(q_{\alpha}-Q_{\alpha}\right)g_{\mu\nu}\,.
\eeq
Finally, upon using the connection field equations (see relation (\ref{fe2}b)) and the fact that $\Delta_{\alpha\mu\nu}= \Delta_{(\alpha\mu)\nu}$, the above expression reduces further to
\beq
N_{\alpha\mu\nu}= -\frac{1}{2}\,\kappa\Delta_{\mu\nu\alpha}+ \frac{1}{2}\,g_{\alpha(\mu}Q_{\nu)}+ \frac{1}{2}\left(q_{\alpha}-Q_{\alpha}\right)g_{\mu\nu}\,.
\eeq

With this in hand and using the post-Riemannian expansions of the Ricci tensor and the Ricci scalar, a lengthy calculation leads to the associated Friedmann equations
\beq
H^{2}= \frac{1}{3}\,\kappa \left[\rho+\frac{1}{2a^{3}}\,\partial_{t}\left(a^{3}\psi\right)\right]+ Y^{2} \hspace{15mm} {\rm and} \hspace{15mm} \frac{\ddot{a}}{a}= -\frac{1}{6}\,\kappa(\rho+3p)+ \frac{1}{a^{3}}\,\partial_{t}\left(a^{3}Y\right)- 2Y^{2}\,,  \label{FRW}
\eeq
where
\beq
Y= -\frac{1}{12}\,\kappa(3\phi+\psi)\,.
\eeq
These relations are supplement by the conservation laws of the matter fields and the non-metricity sources. It can be easily seen that the continuity equation of the matter remains the same, namely
\beq
\dot{\rho}+ 3H(\rho+p)= 0\,.  \label{FRWcont}
\eeq
On the other hand, the evolution of the non-metricity fields $\phi$ and $\psi$, follows form the conservation law of the hypermomentumd. More specifically, taking the trace of Eq.~(\ref{QlawHM}), we obtain
\beq
(3\phi-\psi)^{\cdot} +3H(3\phi-\psi)= 0\,,  \label{p}
\eeq
which implies that $3\phi-\psi\propto a^{-3}$. Furthermore, expanding Eq.~(\ref{QlawHM}) gives
\beq
\delta_{\lambda}{}^{\mu}\partial_{\nu}\left(\sqrt{-g}\phi u^{\nu}\right)+ u_{\lambda}u^{\mu}\partial_{\nu}\left(\sqrt{-g}\,(\phi+\psi) u^{\nu}\right)+ \sqrt{-g}\,(\phi+\psi)(\dot{u}_{\lambda}u^{\mu} +u_{\lambda}\dot{u}^{\mu})= 0\,,
\eeq
the timelike and the spacelike components of which integrate to
\beq
\phi= \phi_0\left(\frac{a_0}{a}\right)^3 \hspace{15mm} {\rm and} \hspace{15mm} \psi= \psi_0\left(\frac{a_0}{a}\right)^3\,,  \label{phipsicls}
\eeq
respectively.\footnote{Taking the time derivative of Eq.~(\ref{FRW}a) and using the conservation laws (\ref{FRWcont}) and (\ref{phipsicls}) leads to (\ref{FRW}b) as expected.} Our last step is to plug these results back into the Friedmann equations (see expressions (\ref{FRW}a) and (\ref{FRW}b) above) and in so doing arrive at
\beq
H^{2}= \frac{1}{3}\,\kappa\rho+ \frac{1}{16}\,\kappa^{2} \left(\phi+\frac{1}{3}\,\psi\right)^{2} \hspace{15mm} {\rm and} \hspace{15mm} \frac{\ddot{a}}{a}= -\frac{1}{6}\,\kappa(\rho+3p)- \frac{1}{8}\,\kappa^{2}\left(\phi+\frac{1}{3}\,\psi\right)^{2}\,.
\eeq
Comparing these results with Eqsd.~($\ref{FriedE2}$a) and ($\ref{FriedE2}$b), we realise that there is no essential differences between the two sets. In fact, one can recover the latter set of formulae from the former by simply introducing the ``shift'' $\phi\rightarrow\phi+\psi/3$ and vice versa. It is therefore not surprising that this type of non-metricity also decelerates the expansion (or accelerates gravitational collapse) and it cannot act as an effective ghost fluid. Nevertheless, this happens for the model of the unconstrained hyperfluid, where the hypermomentum is either of purely dilatonic type, or it is related to it by means of a disformal transformation of the metric.\footnote{Following $(\ref{D2})$, we may write $\Delta_{\alpha\mu\nu}=\tilde{g}_{\alpha\mu}u_{\nu}$, with $\tilde{g}_{\alpha\mu}=\phi g_{\alpha\mu}+ (\phi+\psi)u_{\alpha}u_{\mu}$.} It is possible for non-metric degrees of freedom, associated with more general forms for hypermomentum, to accelerate the expansion, or resist gravitational collpase (see next section and also section~\ref{GNM} below).

\subsection{Weyl non-metricity}
In order to demonstrate the versatility of the non-Riemannian degrees of freedom, let us consider the simple case of Weyl non-metricity, with
\beq
Q_{\alpha\mu\nu}= Au_{\alpha}g_{\mu\nu}\,,  \label{Weyl}
\eeq
with $A=A(t)$. Following a procedure identical to the one outlined above, one arrives at the modified Friedmann equations
\beq
H^{2}=\frac{1}\,\kappa\rho+ HA- \frac{1}{2}\,A^{2} \hspace{15mm} {\rm and} \hspace{15mm} \frac{\ddot{a}}{a}= -\frac{1}{6}\,\kappa(\rho+3p)+ \frac{1}{2}\,\dot{A}+\frac{1}{2}\,HA\,.
\eeq
Accordingly, depending on the signs of $A$ and $\dot{A}$, the hypefluid and act as conventional or ghost-like matter, while at the same time it can either decelerate or accelerate the expansion of the host universe.

Finally, it is worth noting the duality between the above set of Friedmann equations and their purely torsional counterparts obtained in \cite{kranas2019friedmann,Iosifidis:2018jwu}. Indeed, the two sets are mapped to each other upon the duality exchange $A\leftrightarrow-4\phi$.

\section{General non-metricity sourced by the perfect Ccosmological 
hyperfluid}\label{GNM}
The most general form on non-metricity allowed by the high symmetry of the FLRW models, is sourced by the ``perfect cosmological hyperfluid'', recently formulated and developed in~\cite{Iosifidis:2020gth}. As a result, this model opens up a plethora of new phenomenological possibilities when studying the implications of non-metricity for cosmology.

\subsection{The perfect hyperfluid}\label{ssPH=F}
Allowing for the maximum number of non-metric degrees of freedom compatible with the Cosmological Principle, translates into a non-metricity tensor of the form
\beq
Q_{\alpha\mu\nu}= Au_{\alpha}h_{\mu\nu}+ Bh_{\alpha(\mu}u_{\nu)}+ Cu_{\alpha}u_{\mu}u_{\nu}\,, \label{Qnmcos2}
\eeq
spanned by three temporal functions $A$, $B$ and $C$. Recall that  the unconstrained hyperfluid discussed earlier contained only two such parameters. Following~\cite{Iosifidis:2020gth}, the hypermomentum tensor of the perfect hyperfluid reads
\beq
\Delta_{\alpha\mu\nu}= \phi h_{\mu\alpha}u_{\nu}+ \chi h_{\nu\alpha}u_{\mu}+ \psi u_{\alpha}h_{\mu\nu}+ \omega u_{\alpha}u_{\mu} u_{\nu}+ \epsilon_{\alpha\mu\nu\rho}u^{\rho}\zeta\,.  \label{Dform0}
\eeq
The spin part of the above is known to source torsion~\cite{hehl1995metric}. The latter vanishes by default in our study, which transalets into the constraints $\psi=\chi$ and $\zeta=0$. As a result, when dealing with pure non-metricity, the hypermomentum tensor reduces to
\beq
\Delta_{\alpha\mu\nu}=\phi h_{\mu\alpha}u_{\nu}+\psi h_{\nu\alpha}u_{\mu}+\psi u_{\alpha}h_{\mu\nu}+\omega u_{\alpha}u_{\mu} u_{\nu}\,,  \label{Dform1}
\eeq
which is the form we will be using for the rest of this study.

\subsection{Relating non-metricity and hypermomentum}\label{ssRN-MH}
Substituting expression (\ref{Qnmcos2}), together with the associated non-metricity vectors $Q_\mu=(3A-C)u_\mu$ and $q_\mu=[(3B/2)-C]u_\mu$, back into Eq.~(\ref{Palatini}) leads to the (symmetrized) Palatini tensor
\beq
P_{\alpha\mu\nu}= P_{\alpha(\mu\nu)}= \frac{1}{2}\,(A-C)u_{\alpha}h_{\mu\nu}+ \frac{1}{2}\,(B-3A-C)h_{\alpha(\mu}u_{\nu)}- \frac{3}{2}\,Bu_{\alpha}u_{\mu}u_{\nu}\,.
\eeq
The above and Eq.~(\ref{Dform1}) combine in the connection field equation ($\ref{fe2}$b) -- notice the symmetrization in the hypermomentum in the latter expression -- to provide the relations between the non-metricity functions and the sources (i.e. the hypermomentum variables). To be precise, we obtain
\beq
A= -\frac{1}{6}\,\kappa(3\phi+\omega)\,, \hspace{15mm} B= -\frac{2}{3}\,\kappa\omega \hspace{15mm} {\rm and} \hspace{15mm}
C= -\frac{1}{6}\,\kappa(3\phi+\omega+12\psi)\,.  \label{ABC}
\eeq
Alternatively, one arrives at the same result by contracting the connection field equations to express the non-metricity vectors in terms of their hypermomentum counterparts (see Appendix~\ref{ApB}).

We finally note that the case of the unconstrained hyperfluid (see section~\ref{sFUN-M}) is characterised by $\chi=0=\psi=\omega$, which substituted back into Eqs~(\ref{ABC}) give
\beq
C= A= -\frac{1}{2}\,\kappa\phi \hspace{15mm} {\rm and} \hspace{15mm} B= 0\,,
\eeq
in agreement with our earlier results. Therefore, as expected, the unconstrained hyperfluid is a special case of the perfect cosmological pyperfluid.

\subsection{Generalised conservation laws}\label{sGCLs}
The conservation laws of the energy-momentum and the hypermomentum tensors that describe the perfect cosmological hyperfluid follow from Eqs.~(\ref{QlawHM}) and (\ref{QlawEM}) by means of (\ref{Qnmcos2}) and (\ref{Dform1}). Recalling that we have assumed zero torsion ($S_\mu=0$) and we have set $\chi=\psi$, we find
\beq
\dot{\rho}+ 3H(\rho+p)= -\frac{1}{2}\,\psi u^{\mu}u^{\nu}(R_{\mu\nu}+\check{R}_{\mu\nu})  \label{rhop}
\eeq
and
\begin{gather}
-\delta^{\mu}{}_{\lambda}\,\frac{\partial_{\nu}\left(\sqrt{-g}\,\phi u^{\nu}\right)}{\sqrt{-g}}- u^{\mu}u_{\lambda}      \frac{\partial_{\nu}\left[\sqrt{-g}\,(\phi+2\psi+\omega) u^{\nu}\right]}{\sqrt{-g}} +\bigg[\frac{1}{2}\,Q_{\lambda}u^{\mu}- \nabla_{\lambda}u^{\mu}
\nonumber \\
+\left(\frac{1}{2}\,Q^{\mu} -q^{\mu}\right)u_{\lambda} -g^{\mu\nu}\nabla_{\nu}u_{\lambda}\bigg]\psi +2\dot{\psi}u^{\mu}u_{\lambda}- (\phi+2\psi+\omega)(\dot{u}^{\mu}u_{\lambda} +u^{\mu}\dot{u}_{\lambda})= 0\,,  \label{conl2}
\end{gather}
for the matter and the hyperfluid respectively. The former is the modified continuity equation, with non-Riemannian effects (hypermomentum contributions) on the right-hand side, while the latter determines the evolution of the hyperfluid itself.

The conservation law (\ref{conl2}) provides us with two evolution equations for the hypermomentum variables. More specifically, the trace of the above expression yields
\beq
\partial_{\mu}\Big[\Big(3\phi-\omega\Big)u^{\mu}\Big]=0
\eeq
and integrates immediately to give
\beq
3\phi-\omega=\frac{C_{0}}{a^{3}} \label{dil}
\eeq
where $C_{0}=a^3_0(3\phi_0-\omega_0)$ is the integration constant. Moreover, taking the trace of the specelike component of ($\ref{conl2}$), namely setting $\mu=i$, $\lambda=j$ and then contracting along $i$ and $j$, gives
\beq
\dot{\phi}+3H\phi +2H\psi +\frac{1}{6}\,\kappa\psi \Big(3\phi-\omega\Big)= 0\,. \label{claw}
\eeq

Note that the unconstrained hyperfluid model is recovered as the special case when $\psi=0$, as expected. In addition, after computing the right-hand side of the generalized continuity equation given in (\ref{rhop}), we arrive at\footnote{See Appendix~\ref{ApA} for details.}
\beq
\dot{\rho}+ 3H(\rho +p)= -\frac{3}{2}\,\psi \left[\frac{1}{2}\,\dot{B}+\left(A+\frac{1}{2}\,B+C\right)H +{1\over2}\left(\frac{1}{2}\,B-A\right)(C+A) \right]\,. \label{continuity}
\eeq

Before closing we should point out that, in order to completely describe the hypermomentum degrees of freedom, we need to introduce ``equations of state'' between the variables $\phi$, $\psi$ and $\omega$, analogous to that between the pressure ($p$) and the density ($\rho$) of the conventional matter (see section~\ref{ssI} below).

\subsection{Generalised Friedmann equations}\label{ssGFEs}
Starting with the generalised symmetric connection, which differs from its Levi-Civita counterpart due to the non-metricity, one generalises the Ricci tensor and the Ricci scalar. Then, taking the timelike and the spacelike components of the metric field equations, leads to (see Appendix~\ref{ApA})
 \beq
H^{2}= \frac{1}{3}\,\kappa\rho- \frac{1}{2}\,H\left(\frac{3}{2}\,B-A+C\right)- \frac{1}{4}\,\dot{B}+ \frac{1}{8}\,B(A-C)+ \frac{1}{4}\,AC  \label{F1N}
\eeq
and
\beq
\frac{\ddot{a}}{a}= -\frac{1}{6}\,\kappa\left(\rho+3p\right)+ H\left(A+\frac{1}{2}\,C\right)+ \frac{1}{2}\,\dot{A}- \frac{1}{4}\,A(A+C)\,.  \label{F2N}
\eeq
These are the modified Friedmann and Raychaudhuri equations respectively. Together with the conservation laws (\ref{dil})-(\ref{continuity}), the above monitor the evolution of an FLRW-like universe with generalised non-metricity. In what follows, we will attempt to explore the potential implications of our model for cosmology.

\subsection{Closing the system}\label{ssI}
We begin our investigation by employing the relations between the non-metricity and the hypermomentum variables (see Eqs.~(\ref{ABC}) in section~\ref{ssRN-MH} earlier) to express (\ref{continuity}), (\ref{F1N}) and (\ref{F2N}) in terms of $\phi$, $\psi$ and $\omega$. This results into the following expression for the continuity equation
\beq
\dot{\rho}+ 3H(\rho+p)= \frac{1}{2}\,\kappa \left[\dot{\omega}+H\left(3\phi+2\omega+6\psi\right) +\frac{1}{12}\,\kappa\left(3\phi-\omega\right) \left(3\phi+\omega+6\psi\right)\right]\psi\,,  \label{continuity2}
\eeq
supplemented by the generalised Friedmann
\beq
H^2= \frac{1}{3}\,\kappa\rho+ \frac{1}{2}\,\kappa H\left(2\psi+\omega\right)+ \frac{1}{6}\,\kappa\left[\dot{\omega}-\kappa\omega\psi +\frac{1}{24}\,\kappa\left(3\phi+\omega\right) \left(\omega+3\phi+24\psi\right)\right]  \label{F1H}
\eeq
and Raychaudhuri
\beq
\frac{\ddot{a}}{a}= -\frac{1}{6}\,\kappa\left(\rho+3p\right)- \frac{1}{4}\,\kappa H\left(\omega+3\phi+4\psi\right)- \frac{1}{12}\,\kappa\left(3\dot{\phi}+\dot{\omega}\right)- \frac{1}{72}\,\kappa^2\left(3\phi+\omega\right) \left(\omega+3\phi+6\psi\right)\,,  \label{F2H}
\eeq
formulae. The above relations simplify considerably after replacing $\omega$, $\dot{\omega}$ and $\dot{\phi}$ with $\phi$, by means of (\ref{dil}) and (\ref{claw}). Indeed, a straightforward calculation leads to the more transparent expressions\footnote{The system of (\ref{FG}) and (\ref{CL}) is self-consistent, since (\ref{FG}b) can be obtained by taking the time derivative of (\ref{FG}a) and then utilizing the remaining equations.}
\beq
H^2= \frac{1}{3}\,\kappa\rho+ \frac{1}{4}\,{\kappa}^2 \left(\phi-\frac{\mathcal{C}_0}{6a^3}\right)^2\,, \hspace{15mm}
\frac{\ddot{a}}{a}= -\frac{1}{6}\,\kappa\left(\rho+3p\right)- \frac{1}{2}\,{\kappa}^2\left(\phi-\frac{\mathcal{C}_0}{6a^3}\right)^2- \frac{1}{2}\,{\kappa}^2\psi \left(\phi-\frac{\mathcal{C}_0}{3a^3}\right)\,,  \label{FG}
\eeq
\beq
\dot{\rho}+ 3H(\rho+p)= \frac{1}{2}\,\kappa\left[H+\frac{1}{2}\,\kappa
\left(\phi-\frac{\mathcal{C}_0}{6a^3}\right)\right]\psi \frac{\mathcal{C}_0}{a^3} \hspace{15mm} \text{and} \hspace{15mm}
\dot{\phi}+ H(3\phi+2\psi)+ \kappa\psi\,\frac{\mathcal{C}_0}{6a^3}= 0\,.  \label{CL}
\eeq
In (\ref{FG}a) one can now see that the extra term due to non-metricity is always positive, which ensures that the hypermomentum acts as a non-ghost fluid with a positive effective energy density. In the second of the Friedmann equations (namely in Raychaudhuri's formula -- see (\ref{FG}b)), the first of the two extra terms always decelerates the expansion. The second non-metricity term, on the other hand, has no definite sign and safe conclusions can be drawn only after solving the full system of equations. However, in order to do so, we need to take additional steps.

Let us turn or attention to the modified continuity equation (\ref{CL}a), which has nonzero right-hand side due to the non-metricity effects. The latter propagate via $\phi$ and $\psi$, of which only $\phi$ has an evolution law (given by (\ref{CL}b) above). Therefore, to proceed, we need to introduce an effective ``equation of state'' of the form $\psi=\psi(\phi)$. Such a step is physically motivated by our desire to represent the hypermomentum as an effective fluid. Then, given that only $\phi$ appears in Eq.~(\ref{FG}a) and $\psi$ appears only in (\ref{FG}b), we may associate the former with the effective energy density and the latter with the effective pressure. On these grounds, the simplest equation of state to assume is a ``barotropic'' one with $\psi=w_1\phi$.\footnote{This is one of the many possibilities, since hyperfluids could generally obey more ``exotic'' equations  of state. As we will see, the simple cases presented below will naturally emerge as subcases of such a ``barotropic'' hyperfluid.} In our case, it is convenient to introduce the relation
\beq
\psi=w_1\Phi\,,
\eeq
with
\beq
\Phi=\phi-\frac{C_0}{6a^3}\,,
\eeq
is a combination that appears in all three of (\ref{FG}a), (\ref{FG}b) and (\ref{CL}a). On using the above equation of state, the system (\ref{FG}a)-(\ref{CL}b) recasts into
\beq
H^2= \frac{1}{3}\,\kappa\rho+ \frac{1}{4}\,\kappa^2\Phi^2\,, \hspace{15mm}
\frac{\ddot{a}}{a}= -\frac{1}{6}\,\kappa(\rho+3p)- \frac{1}{2}\,\kappa^2\Phi^2- \frac{1}{2}\,\kappa^2w_1\Phi \left(\Phi-\frac{C_0}{6a^3}\right)\,,  \label{FGE}
\eeq
\beq
\dot{\rho}+ 3H(\rho+p)= \frac{1}{2}\,\kappa w_1\Phi \left(H+\frac{1}{2}\,\kappa\Phi\right)\frac{C_0}{a^3} \hspace{15mm} {\rm and} \hspace{15mm} \dot{\Phi}+ (3+2w_1)H\Phi+ \frac{1}{6}\,\kappa w_1\Phi\,\frac{C_0}{a^3}= 0\,,  \label{CL2}
\eeq
where now $C_0=2a_0^3(3\Phi_0-\omega_0)$. Given that the last term in (\ref{FGE}b) has no definitive sign, one may in principle select an equation of state that could drive cosmic acceleration. Setting $w_1=-1$ and demanding that $C_0\Phi>0$, for example, ensures that the last term in the modified Raychaudhuri equation tends to accelerate the expansion.

\subsection{Solutions}
Let us now turn our attention to specific scenarios that also represent physically motivated simplifications of the general case.
The hypermomentum tensor splits into its irreducible parts, namely spin, dilation and shear as follows~\cite{hehl1995metric}
\beq
\Delta_{\alpha\mu\nu}= \Delta_{[\alpha\mu]\nu}+\frac{1}{4}\,g_{\alpha\mu}\mathit{D}_\nu+ \breve{\Delta}_{\alpha\mu\nu}\,.
\eeq
Given that the spin part (i.e.~first term) is due to torsion, we will hereafter exclude it from our analysis. Of the remaining terms, the dilation part is equal to
\beq
\mathit{D}_\nu:= \Delta_{\alpha\mu\nu}g^{\alpha\mu}= (3\phi-\omega)u_\nu\,,  \label{dilation}
\eeq
while the symmetric traceless shear part reads
\beq
\breve{\Delta}_{\alpha\mu\nu}= 2\psi u_{[\mu}h_{\alpha]\nu}+ \frac{1}{4}\,(\phi+\omega)h_{\alpha\mu}u_\nu+ \frac{3}{4}\,(\phi+\omega)u_\nu u_\mu u_\alpha\,.  \label{shear}
\eeq
In what follows, we isolate the effects of these two sources and investigate their potential cosmological implications separately.

\subsubsection{Pure dilation}
Let us assume that the shear component vanishes and that only the dilation part of the hypermomentum survives. As seen from Eq.~($\ref{shear})$, this leads to the constraints
\beq
\breve{\Delta}_{\alpha\mu\nu}=0 \Leftrightarrow \left\{ \begin{array}{rl}  \psi & =0 \\ \phi+\omega & =0 \end{array} \right.\,.
\eeq
Setting $\psi=0$ simplifies our system of equations considerably, while applying the $\phi+\omega=0$ condition to Eq.~($\ref{dil}$), the latter gives
\beq
 \Phi=\frac{C_0}{12a^3}\,,
\eeq
which follows form ($\ref{CL2}$) as well. When $\psi=0$, the continuity equation does not explicitly depend on the hypermomentum. Then, supplementing a barotropic equation of state ($p=w\rho$) for the matter leads to $\rho\propto a^{-3(1+w)}$. On these grounds, the Friedmann and the Raychaudhuri equations read
\beq
H^2=\frac{1}{3}\,\kappa\rho+ \kappa^2\frac{C_0^2}{24^2a^6}
\hspace{15mm} {\rm and} \hspace{15mm}
\frac{\ddot{a}}{a}= -\frac{1}{6}\,\kappa(1+3w)\rho- \frac{1}{2}\,\kappa^2\frac{C_0^2}{12^2a^6}\,,
\eeq
respectively. Therefore, qualitatively speaking, the effective energy density contribution of the pure-dilation case is always positive, which means that the extra non-metricity term in the Raychaudhuri equation is negative and decelerates the expansion at all times (just like conventional matter with $1+3w>0$). To facilitate the presentation, let us define the effective energy density for the hyperfluid as
\beq
\rho_e= \frac{1}{3}\,\kappa\frac{C_0^2}{4^3a^6}  \label{Phidil}
\eeq
and recast the above into
\beq
H^2=\frac{1}{3}\,\kappa(\rho+\rho_e) \hspace{15mm} {\rm and} \hspace{15mm} \frac{\ddot{a}}{a}= -\frac{1}{6}\,\kappa(1+3w)\rho- \frac{2}{3}\,\kappa\rho_e\,.  \label{FEDil}
\eeq
Accordingly, in dynamical terms, non-metricity mimics a classical stiff fluid, namely like a perfect medium with barotropic index $w=1$. This behaviour is essentially identical to the one found in sections \textit{III} and \textit{IV} earlier. Overall, generalising the host spacetime to include non-metricity, provides an alternative (geometrical) way of producing an effective stiff component.

From the stiff matter interpretation of pure-dilation non-metricity it becomes evident that, even if non-metricity is dynamically negligible today, it could have dominated at sufficiently early times and thus it might have dictated the kinematical evolution of the universe in the past. Indeed, Eq.~(\ref{FEDil}a) accepts the solutions
\begin{equation}
a\propto t^{1/3} \hspace{15mm} {\rm and} \hspace{15mm} a\propto t^{2/3(1+w)}\,, \label{mFsol1}
\end{equation}
at early and late times respectively. Note that the first result corresponds to $\rho_e\gg\rho$ and the latter to $\rho_e\ll\rho$. Also, the two solutions coincide when $w=1$, namely when the conventional matter also has a stiff equation of state. The possibility raised by solution (\ref{mFsol1}), is intriguing, since it provides theoretical support to speculations that matter with hyperfluid characteristics could have dominated the very early stages of the expansion. Here, we see that this can happen naturally within the framework of MAG, by allowing for non-Riemannian geometry (i.e.~for spacetime non-metricity). Having said that, assuming that non-metricity dominates, the timing is crucial because it could interfere with fundamental physical processes, such as nucleosynthesis for example (see next section).

Let us now consider analytical solutions for specific values of the barotropic indices. Setting $w=1=w_e$, for example, the two fluids have identical behaviour. The solution is the identical to its standard GR counterpart, with the matter density replaced by the sum of sum of the densities of the fluids
\beq
a=a_0 \left(\sqrt{\frac{3}{4}\,\kappa \left(\rho_0+\rho_{e_0}\right)}(1+w)t\right)^{1/3}\,,
\eeq
where the zero suffix denotes the present time. Assuming that matter is in the form of pressure-free dust (i.e.~setting $w=0$) leads to
\beq
a=a_0 \left(\sqrt{\frac{3}{4}\,\kappa\rho_{e_0}}t +\frac{3}{4}\,\kappa\rho_0 t^2\right)^{1/3}\,.
\eeq
The first mode in the above grows as $\propto t^{1/3}$ and dictates the early evolution of the model, while the second is proportional to $t^2/3$ and drives the late-time expansion (both in agreement with solutions (\ref{mFsol1})).

Let us also consider the effects of pure-dilation non-metricity on standard inflation. Assuming matter with a de Sitter inflationary equation of state, we may set $w=-1$. Then, we obtain the solution
\beq
a=a_0\left(\frac{\rho_{e}}{\rho}\right)_0^{1/6} \left[\sinh{\left(\sqrt{3\kappa\rho_0}t\right)}\right]^{1/3}\,,
\eeq
which also depends on the ratio $\rho_{e}/\rho$ today. At early times, that is for small values of $t$, the above approaches solution (\ref{mFsol1}a) and therefore inflation is suppressed. In contrast, at late times, we recover the exponential expansion of standard de Sitter-type inflation.

\subsubsection{Pure shear}\label{sssPS}
Turning our attention to pure shear hypermomentum, we set the dilation component of $\Delta_{\alpha\mu\nu}$ (see Eq.~(\ref{dilation})) to zero. Combining the latter with ($\ref{dil}$), gives
\beq
\mathit{D}_\nu= 0 \Leftrightarrow C_0= 0
\eeq
which in turn recast the system of (\ref{FGE}) and (\ref{CL2}) into
\beq
H^2= \frac{1}{3}\,\kappa\rho+ \frac{1}{4}\,\kappa^2\Phi^2\,,
\hspace{25mm}
\frac{\ddot{a}}{a}= -\frac{1}{6}\,\kappa(\rho+3p)- {1\over2}\,\kappa(1+w_1)\Phi^2\,,  \label{modFRW}
\eeq
\beq
\dot{\rho}+ 3H(\rho+p)= 0 \hspace{15mm} {\rm and} \hspace{15mm}
\dot{\Phi}+ (3+2w_1)\Phi H= 0\,, \label{CLshear}
\eeq
respectively. As it turns out, the above relations are the same with the ones derived in~\cite{Iosifidis:2020upr}, were both torsion and non-metricity were included, though there the hypermomentum tensor had only shear component. For completeness, we will briefly recover some of the solutions given in~\cite{Iosifidis:2020upr}, before providing additional results. Following (\ref{modFRW}), the non-metricity input to the generalized Friedmann equation is always positive, while its contribution to the Raychaudhuri equation depends on $w_1$-index. In addition, according to (\ref{CLshear}), the two continuity equations have decoupled, which guarantees that matter evolves independently of the $\Phi$-field. The physical interpretation becomes clearer if we introduce the effective energy density
\beq
\rho_e= \frac{3}{4}\,\kappa\Phi^2 \,.
\eeq
On using the above, the associated onservation law (see Eq.~(\ref{CLshear}b)) becomes identical to the continuity equation of a barotropic perfect fluid, namely
\beq
\dot{\rho_e}+3H(1+w_e)\rho_e=0\,, \label{modCL}
\eeq
with $w_e=[1+(4w_1/3)]$. At the same time, the Friedmann and the Raychaudhuri equations given in (\ref{modFRW}) read
\beq
H^2=\frac{1}{3}\,\kappa\rho+ \frac{1}{3}\,\kappa\rho_e \hspace{15mm} {\rm and} \hspace{15mm}
\frac{\ddot{a}}{a}= -\frac{1}{6}\,\kappa(1+3w)\rho- \frac{1}{6}\,\kappa(1+3w_e)\rho_e\,,  \label{modFRW1}
\eeq
respectively. This cosmology is described by usual matter plus an additional effective perfect fluid that has emerged after generalizing the Riemannian geometry of the host spacetime by including non-metricity effects. Following (\ref{modFRW1}a) conventional matter dominates when $\rho\gg\rho_e$. Otherwise, it is the non-metricity that dictates the universal kinematics. According to Eq.~(\ref{modFRW1}b), matter that satisfies the strong energy condition (i.e.~with $1+3w>0$) decelerates the expansion (as expected). Similarly, the (hypermomentum) second term on the right-hand side of (\ref{modFRW1}b) slows the expansion down when $1+3w_e>0$, or equivalently when $w_1>-1$. In the opposite case the expansion is accelerated, whereas setting $w=-1/3=w_e$ leads to $q=0$ and to the familiar ``coasting universe''. We should also point out that, at the moment, there are no known constraints imposed on $w_1$ and therefore on $w_e$.

The continuity equations of the conventional and the effective fluids (see expressions (\ref{CLshear}a) and (\ref{modCL}) previously) are integrated immediately, respectively giving
\beq
\rho= \rho_0\Big(\frac{a_0}{a}\Big) ^{3(1+w)} \hspace{15mm} {\rm and} \hspace{15mm} \rho_e= \rho_{e_0} \Big(\frac{a_0}{a}\Big)^{3(1+w_e)}\,,
\eeq
where $\rho_0$ and $(\rho_e)_0$ are the matter density and the effective density of the hypermomentum component today. On using the above, the generalized Friedmann equation (see expression (\ref{modFRW1}a)) yields
\beq
H^2= \frac{1}{3}\,\kappa\rho_0\Big(\frac{a_0}{a}\Big)^{3(1+w)} \bigg[1+\Big(\frac{\rho_e}{\rho}\Big)_0 \Big(\frac{a_0}{a}\Big)^{3(w_e-w)}\bigg]\,.  \label{modH2}
\eeq
The above implies that, although the hypermomentum component may be subdominant today (i.e.~even if $(\rho_e)_0\ll\rho_0$), non-metricity could have dominated in the distant past, or it may take over in the far future. Which of these alternative scenarios occurs depends on the sign of the difference $w_e-w$ between the two barotropic indices. More specifically, when $w_e-w>0$, the shear hypermomentum dominates the early evolution of the universe. In the opposite case, that is for $w_e-w<0$, this happens in the late future. Clearly, when $w_e=w$, a universe that starts matter-dominated remains so at all times and vice versa.

Provided that $w_e-w\neq0$, the two constituents reach equipartition, analogous to that between radiation and dust, at a specific moment in the lifetime of the universe, which can lie either in the past or in the future. Following (\ref{modH2}), the scale factor at the equilibrium point is
\beq
a_{\star}= a_0 \left(\frac{\rho_e}{\rho}\right)_0^{[1/3(w_e-w)]}\,.
\eeq
Making the plausible assumption that $\rho_e/\rho\ll1$ at present, we may set $(\rho_{e}/\rho)_0=10^{-\beta}$, with $\beta$ being a small positive number. Then, the scale factor at equilibrium will be
\beq
a_{\star}= a_0\times10^{-[\beta/3(w_e-w)]}\,.
\eeq
The above ensures that $a_{\star}<a_0$ when $w_e-w>0$, which puts the time of equipartition in the past. In the opposite case we find that $a_{\star}>a_0$ and non-metricity takes over in the future. Finally, recalling that the temperature of the universe evolves inversely proportional to the scale factor, namely that $T\propto 1/a$, the temperature at the moment of equipartition is given by
\beq
T_{\star}=T_0\times10^{[\beta/3(w_e-w)]}\,,
\eeq
with $\beta>0$. Clearly, $T_{\star}>T_0$ when $w_e-w>0$ and $T_{\star}<T_0$ for $w_e-w<0$, putting the equilibrium time in the past and in the future respectively.

\section{Conclusions}
In this work we have examined the role of spacetime non-metricity in homogeneous and isotropic (Friedmann-like) cosmologies. Non-metricity reflects the failure of the metric to be covariantly conserved, which changes the magnitude of vectors and tensors when they are parallelly transported. Many familiar geometric notions change in the presence of non-metricity, such as the symmetries of the Riemann tensor, its contractions, the conservation laws, etc. It is therefore not surprising that including non-metricity changes the cosmic evolution in non-trivial ways.

We begun our investigation by considering the usual Einstein-Hilbert action and a matter sector that also depends on the connection, namely with non-vanishing hypermomentum. The spacetime was assumed torsionless to begin with, that is the connection was symmetric by default, in order to isolate the effects of non-metricity. Within this framework, we provided the connection, the field equations and the associated conservation laws. Turning our attention to cosmology, we considered a spatially flat Fridemann-like universe along with the allowed form for the distortion and non-metricity tensors. Clearly, the symmetries of the FRLW metric restrict the allowed forms of the various energy tensors. More specifically, the energy-momentum tensor can only have the perfect fluid form and the hypermomentum must be compatible;e with the Cosmological Principle~\cite{Iosifidis:2020gth}.

With this in hand, we examined the unconstrained hyperfluid developed in~\cite{obukhov1996model}. This model can source two of the three non-metricity degrees of freedom allowed in a cosmological setting. Nevertheless, having extracted the modified Friedmann equations and after employing the associated cosmological laws, we obtained a number of results. The conservation laws of the energy-momentum and the hypermomentum tensors, in particular, decouple and can be trivially integrated to give the evolution laws (i.e.~the continuity equations) of $\rho$ , as well as those of $\phi$ and $\psi$. The latter showed that both $\psi$ and $\phi$ evolve like pressureless matter, with $\phi,\,\psi\propto1/a^{3}$, where $a=a(t)$ is the cosmological scale factor). Moreover, the fact that the aforementioned two fields contribute quadratically to the modified Friedmann equations implies that their input to the cosmological dynamics is proportional to $1/a^{6}$. Consequently,    the effect of $\phi$ and $\psi$ is phenomenologically identical to that of a perfect fluid with a stiff equation of state. Put another way, an effective stiff-matter component emerges naturally as a cosmological consequence of the underlying non-Riemannian geometry of the universe and without any further assumptions.

We have also explored in depth the most general form of non-metricity sourced by a perfect hypermomentum-preserving hyperfluid~\cite{Iosifidis:2020gth}. This generalized model is described by three additional degrees of freedom, both for the non-metricity and the hypermomentum that sources it. The complete system of equations immediately led to some qualitative conclusions. For example, the additional terms introduced to the Friedmann equation were found to be positive always, indicating that non-metricity mimics matter with positive (i.e.~non-ghost-like) effective energy density. The extra terms in the modified Raychaudhuri equation, on the other hand, suggested that non-metricity can accelerate or decelerate the universal expansion, depending on the (effective) equation of state one assumes for the hypermomentum variables.

Looking for analytic solutions, we examined two physically distinct subcases, that is for hypermomentum with a dilation part only and for shear hypermomentum. In the former case, we recover the (effective) stiff-matter scenario mentioned above and also provided exact cosmological solutions, corresponding to specific values of the barotropic index ($w$) of the matter. For shear hypermomentum, we found that the effects of non-metricity mimic those of a perfect fluid with a ``free'' barotropic index ($w_e$). This freedom provides  a rich phenomenology, since different values of $w_e$ can lead to universal acceleration, or deceleration, to exponential expansion, etc. In this latter scenario, there is also a distinct moment in time, where the conventional matter and the non-metricity induced (effective) fluid have equal energy densities (i.e.~$\rho_e=\rho$). Such a ``moment of equipartition'' is dynamically similar to the ``equilibrium point'', marking the transition from the radiation to the dust era of the universe. In fact, assuming that the current value of $\rho_{e}$ can be constrained, one could determine the transition time from a non-metricity dominated epoch of the universe to the era of conventional matter domination (and vice versa). Depending  on the sign of the difference $w-w_e$, this moment can lie in our past or in our future.

Before closing, we should point out that our initial assumption that the canonical and the metrical energy momentum tensors coincide, restricts the allowed forms of non-metricity. For example, ``fixed length non-metricity'' is incompatible with our equations.\footnote{Where vector lengths are preserved  and the totally symmetric part of the non-metricity tensor is identically zero} In order to include more general types of non-metricity, one needs to ease the restrictions imposed in Eq.~($\ref{conl2}$) and thus break away from the hypermomentum-conserving hyperfluids~\cite{Iosifidis:2021fnq}. We plan consider such generalized non-metricity models in the near future.

\newpage

\appendix

\section{Curvature in the presence of general 
non-metricity}\label{ApA}
Here we provide the curvature tensors and scalars necessary for the derivation of the modified Friedman equations and the associated conservation laws. Following section~\ref{ssCC}, the general affine connection splits into its Riemannian and Non-Riemannian components as $\Gamma^\lambda{}_{\mu\nu}=\tilde{\Gamma}^\lambda{}_{\mu\nu}+ N^\lambda{}_{\mu\nu}$, with
 \beq
N^\lambda{}_{\mu\nu}= \frac{1}{2}\,g^{\rho\lambda} \left(Q_{\mu\nu\rho}+Q_{\nu\rho\mu}-Q_{\rho\mu\nu}\right)\,.
\eeq
being the distortion tensor in the absence of torsion. Splitting the connection leads to the decomposition of the Riemann tensor and its contractions. In particular,
\beq
R^\mu{}_{\nu\alpha\beta}= \tilde{R}^\mu{}_{\nu\alpha\beta}+ 2\tilde{\nabla}_{[\alpha}N^\mu{}_{|\nu|\beta]}+ 2N^\mu{}_{\rho[\alpha}N^\rho{}_{|\nu|\beta]}\,, \label{RiemannTensorDecompositionA}
\eeq
with the tilded quantities evaluated with respect to the Levi-Civita connection. For the most general form of non-metricity compatible with the cosmological principle, namely when $Q_{\alpha\mu\nu}= Au_{\alpha}h_{\mu\nu}+Bh_{\alpha(\mu}u_{\nu)}+ Cu_{\alpha}u_{\mu}u_{\nu}$, the distortion tensor becomes
\beq
N^\lambda{}_{\mu\nu}= \frac{1}{2}\,(B-A)h_{\mu\nu}u^\lambda+ A\delta^\lambda{}_{(\mu}u_{\nu)}+\left(A+\frac{1}{2}\,C\right )u^\lambda u_\mu u_\nu\,.  \label{distortionTensorFRWA}
\eeq

Contracting (\ref{RiemannTensorDecompositionA}) and using the above distortion tensor, as well as its contractions, leads to the curvature variables. The Ricci tensor, in particular, reads
\begin{eqnarray}
R_{\nu\beta}&=& \tilde{R}_{\nu\beta}+ \frac{1}{2}\,g_{\nu\beta}\tilde{\nabla}_\mu\left[(B-A)u^\mu\right]+ \frac{1}{2}\,\tilde{\nabla}_\beta \left[\left(C-2A\right)u_\nu\right]+ \frac{1}{2}\,\tilde{\nabla}_\nu \left(Au_\beta\right)+ \frac{1}{2}\,\tilde{\nabla}_\mu \left[(A+B+C)u_\nu u_\beta u^\mu\right]\nonumber\\
&&+\frac{1}{4}\,(A-B)(A-C)h_{\nu\beta}- \frac{3}{4}\,A(A+C)u_\nu u_\beta\,.  \label{RicciFRWFinalA}
\end{eqnarray}
Then, after taking into account the FLRW metric and the associated the Levi-Civita connection, we obtain
\beq
R_{00}= \tilde{R}_{00}+ \frac{3}{2}\,\dot{A}+ \frac{3}{2}\,(2A+C)H- \frac{3}{4}\,A(A+C)
\eeq
and
\beq
R_{ii}= \tilde{R}_{ii}+ \frac{1}{2}\,(C+3B-4A)a\dot{a}+ \frac{1}{2}\,a^2(\dot{B}- \dot{A})+ \frac{1}{4}\,a^2(B-A)(C-A)\,,
\eeq
for the timelike and spacelike compontents respectively. Contracting again, leads to the Ricci scalar
\beq
R= \tilde{R}+ \frac{3}{2}\,\dot{B}- 3\dot{A}+ 9H\left(\frac{1}{2}\,B-A\right)+ \frac{3}{2}\,A^2+ \frac{3}{4}\,B(C-A)\,.
\eeq
These results combine with the Einstein field equations to give the generalised Friedmann equations presented in section~\ref{ssGFEs} of the main document.

We also note that the Co-Ricci tensor involved in the conservation law of the hypermomentum reads
\begin{eqnarray}
\check{R}_{\lambda\beta}&=& -\tilde{R}_{\lambda\beta}+ \frac{1}{2}\,\tilde{\nabla}_\beta\left[(2A-2B+C)u_\lambda\right]+ \frac{1}{2}\,g_{\lambda\beta}\tilde{\nabla}_\alpha \left(Au^\alpha\right)+ \frac{1}{2}\,\tilde{\nabla}_\lambda \left(Au_\beta\right)+ \frac{1}{2}\,\tilde{\nabla}_\mu \left[(A+B+C)u_\lambda u_\beta u^\mu\right]\nonumber\\
&&-\frac{1}{4}\,A(3A-2B+C)h_{\lambda\beta}+ \frac{3}{4}\,(B-A)(A+C)u_\lambda u_\beta\,.  \label{CoRicciFRWfinalA}
\end{eqnarray}
Finally, the homothetic curvature is antisymmetric and thus always zero in isotropic spaces.

\section{Alternative derivation of the non-metricity tensor and 
vectors}\label{ApB}
One can arrive at Eqs.~(\ref{ABC}) given in section~\ref{ssRN-MH} by taking contractions of the connection field equations to express the non-metricity vectors in terms of those of the hypermomentum. The results read
\beq
Q^{\lambda}=\kappa \left[\Delta^{\lambda\alpha}{}_{\alpha} -\frac{1}{3}\,(\Delta_{\alpha}{}^{\alpha\lambda} +\Delta_{\alpha}{}^{\lambda\alpha})\right]
\hspace{15mm} {\rm and} \hspace{15mm}
q^{\lambda}= \frac{1}{2}\,\kappa \left[\Delta^{\lambda\alpha}{}_{\alpha} +\frac{1}{3}\,\left(\Delta_{\alpha}{}^{\alpha\lambda} +\Delta_{\alpha}{}^{\lambda\alpha}\right)\right]\,.
\eeq
Plugging the above back into the connection field equations and solving for the non-metricity tensor, we arrive at
\beq
Q_{\lambda}{}^{\mu\nu}= \kappa\left(-\Delta_{\lambda}{}^{(\mu\nu)} +\frac{2}{3}\,\Delta_{\alpha}{}^{\alpha(\mu}\delta^{\nu)}_{\lambda} +D_{\lambda}g^{\mu\nu}\right)\,,
\eeq
where
\beq
D_{\lambda}= \frac{1}{2}\left(\Delta_{\lambda\alpha}{}^{\alpha} -\frac{2}{3}\,\Delta^{\alpha}{}_{\alpha\lambda}\right)\,.
\eeq
This procedure also leads to relations (\ref{ABC}) between the two sets of variables.

\bibliographystyle{unsrt}
\bibliography{ref}

\end{document}